\documentclass[aip,amsmath,amssymb,reprint,graphicx]{revtex4-1}
\usepackage{graphicx}
\usepackage{dcolumn}
\usepackage{bm}
\usepackage{float}

\usepackage[utf8]{inputenc}
\usepackage[T1]{fontenc}
\usepackage{mathptmx}
\usepackage{siunitx}  
\usepackage{amsmath}
\usepackage{subfigure}
\usepackage[hidelinks]{hyperref}
\usepackage[all]{hypcap}
\usepackage[noabbrev, capitalise]{cleveref}
\usepackage{amssymb}

\usepackage{natbib}

\draft

\begin{document}

\preprint{AIP/123-QED}

\title{Optomechanical methodology for characterizing the thermal properties of 2D materials}
\author{Hanqing Liu}
\affiliation{Department of Precision and Microsystems Engineering, Delft University of Technology, Lorentzweg 1, 2628 CD Delft, The Netherlands.}

\author{Hatem Brahmi}
\affiliation{ASML Netherlands, B.V., 5504 DR, Veldhoven, The Netherlands.}

\author{Carla Boix-Constant}
\affiliation {Instituto de Ciencia Molecular (ICMol), Universitat de Valencia, Paterna 46980, Spain.}

\author{Herre S. J. van der Zant}
\affiliation{Kavli Institute of Nanoscience, Delft University of Technology, 2628 CJ Delft, The Netherlands.}

\author{Peter G. Steeneken}
\affiliation{Department of Precision and Microsystems Engineering, Delft University of Technology, Lorentzweg 1, 2628 CD Delft, The Netherlands.}
\affiliation{Kavli Institute of Nanoscience, Delft University of Technology, 2628 CJ Delft, The Netherlands.}

\author{Gerard J. Verbiest}
    
\affiliation{Department of Precision and Microsystems Engineering, Delft University of Technology, Lorentzweg 1, 2628 CD Delft, The Netherlands.}

\email{G.J.Verbiest@tudelft.nl}
\email{H.Liu-7@tudelft.nl}

\date{\today}

\begin{abstract}
Heat transport in two-dimensions is fundamentally different from that in three dimensions. As a consequence, the thermal properties of 2D materials are of great interest, both from scientific and application point of view. However, few techniques are available for accurate determination of these properties in ultrathin suspended membranes. Here, we present an optomechanical methodology for extracting the thermal expansion coefficient, specific heat and thermal conductivity of ultrathin membranes made of 2H-TaS$_2$, FePS$_3$, polycrystalline silicon, MoS$_2$ and WSe$_2$. The obtained thermal properties are in good agreement with values reported in the literature for the same materials. Our work provides an optomechanical method for determining thermal properties of ultrathin suspended membranes, that are difficult to measure otherwise. It can does provide a route towards improving our understanding of heat transport in the 2D limit and facilitates engineering of 2D structures with dedicated thermal performance.
\end{abstract}

\maketitle

\section{Introduction}
Soon after the discovery of monolayer graphene, it was found that 2D materials have unique thermal properties, which open opportunities for heat control at the nanoscale \cite{wang2017thermal,mas20112d,wu2022high,gu2016phonon,pop2012thermal}. Due to their ultrasmall thickness, thermal properties of 2D materials are dominated by surface scattering of acoustic phonons, which is highly sensitive to strain \cite{liu2016disparate}, grain size \cite{ying2019tailoring} and temperature \cite{luo2015anisotropic}, as well as material imperfections such as defects and impurities \cite{gu2018colloquium}. To understand and optimize heat transport in 2D materials, precise thermal characterization methods are of great importance.

So far, a variety of experimental techniques have been developed to characterize thermal transport in 2D materials, of which the transient micro-bridge method \cite{jo2014basal,wang2016superior} and the steady-state optothermal method based on Raman microscopy are most commonly used \cite{balandin2008superior,zhou2014high}. However, the construction of a micro-bridge is complicated and thermal contact resistances can affect measurement results, while for Raman measurements, the probed temperature resolution is usually relatively small, leading to large error bars. These limitations undermine the accuracy of probing heat transport in 2D materials, causing large variations in the thermal material parameters reported in literature. For example, literature values for the thermal conductivity vary from 2000 to 5000~\si{Wm^{-1}K^{-1}} for suspended monolayer graphene \cite{nika2017phonons}.

In this work, we demonstrate an optomechanical non-contact method for measuring the thermal properties of nanomechanical resonators made of free-standing 2D materials. The presented methodology allows us to simultaneously extract the thermal expansion coefficient, the specific heat and the in-plane thermal conductivity of the material. It involves driving a suspended membrane using a power-modulated laser and measuring its time-dependent deflection with a second laser. Thus both the temperature-dependent mechanical fundamental resonance frequency of the membrane and characteristic thermal time constant at which the membrane cools down \cite{dolleman2017optomechanics} are measured. A major advantage of the method is that no physical contact needs to be made to the membrane, such that its pristine properties are probed and no complex device fabrication is needed. Buckling effects are incorporated in the model to account for the induced compressive stress during temperature variations. Our results on 2H-TaS$_2$, FePS$_3$, polycrystalline silicon (Poly Si), MoS$_2$ and WSe$_2$ show good agreement with reported values in the literature. 

\begin{figure*}
	\centering  \includegraphics[width=0.7\linewidth,angle=0]{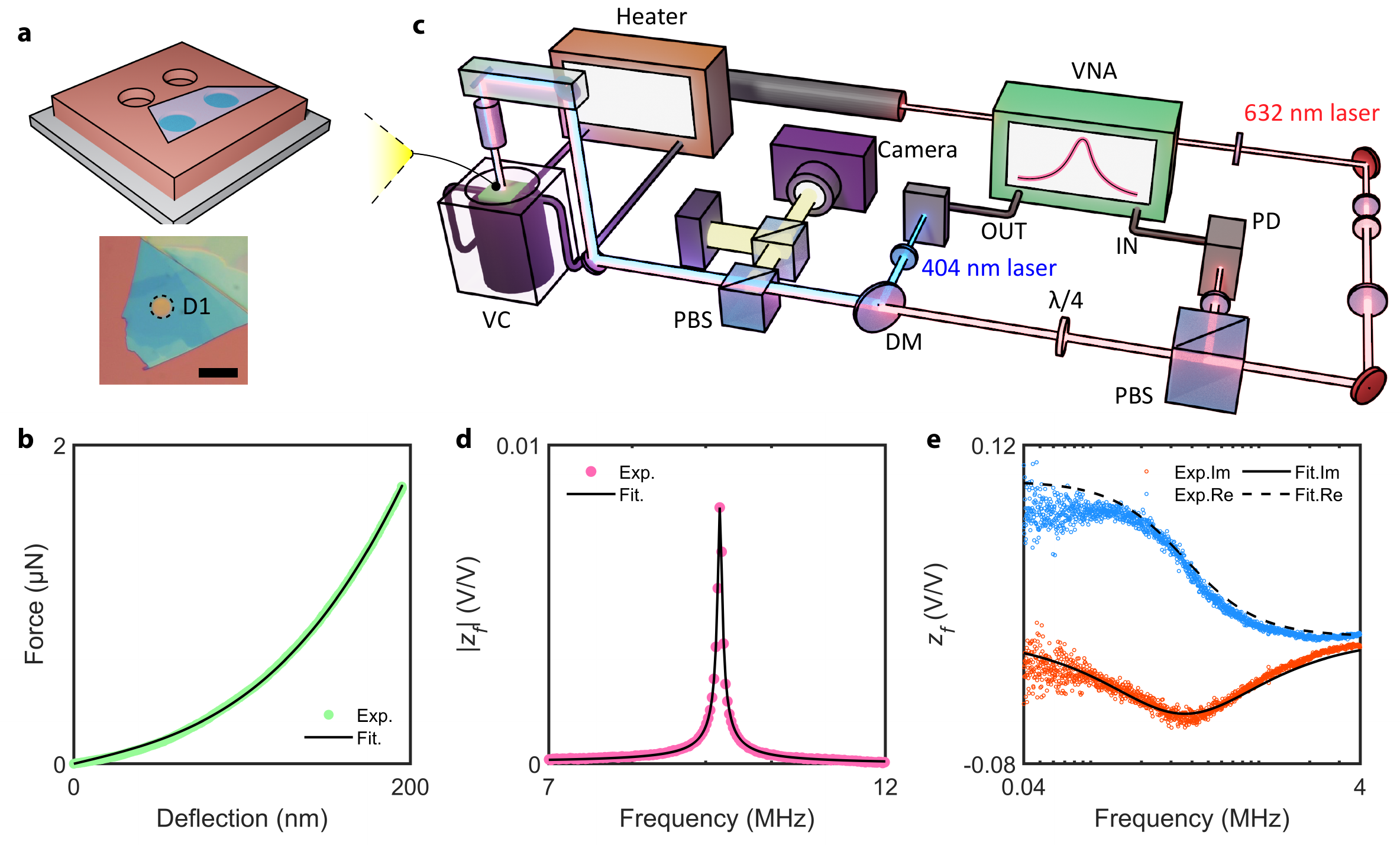}
	\caption{Sample characterization and experimental setup. \textbf{a} Top, schematic diagram of 2D nanomechanical resonators, composed of 2D flake suspended on the etched SiO$_2$/Si cavities; bottom, optical image of device D1 (2H-TaS$_2$). Scale bar is 5~\si{\mu m}. \textbf{b} AFM indentation results for device D1 (points), from which the Young's modulus $E$ of the membrane is extracted by fitting the measured force $F$ to the cantilever deflection $\delta$ (drawn line). \textbf{c} Laser interferometry setup used for the optomechanical measurement. The sample is mounted in a vacuum chamber (VC) with a pressure below 10$^{-5}$~mbar. The reflected red laser is detected by the photodiode (PD) and input to the vector network analyzer (VNA). PBS, polarized beam splitter; DM, Dirac mirror. \textbf{d} Resonant peak of device D1 measured at MHz regime (points), which is fitted with a harmonic model (drawn line) to extract the fundamental  resonance frequency $f_0$ of device D1. \textbf{e} Thermal signal measured at kHz regime, including imaginary (red points) and real (blue points) parts. The imaginary part is fitted with Eq.~(\ref{eq:1}) (drawn lines) to obtain the thermal time constant $\tau$ of device D1.}
	\label{fig:1}
\end{figure*}

\section{Fabrication and methodology}

We fabricate 2D nanomechanical resonators by transferring 2D flakes over circular cavities with a depth of $285$~\si{\nano\meter} and a radius $R$ of 3 to 4~\si{\mu m} in a silicon (Si) substrate  with a 285~\si{nm} thick silicon oxide (SiO$_2$) layer, as illustrated in Fig.~\ref{fig:1}a. The devices D1$-$D5 studied in this work are made of 2H-TaS$_2$, FePS$_3$, Poly Si, MoS$_2$ and WSe$_2$, respectively. By using tapping mode Atomic Force Microscope (AFM), we determine the thickness, $h$, of each membrane (see Table~\ref{tab:op_table1}). All details about the device fabrication and thickness measurement can be found in SI section 1. To determine the Young's modulus $E$ of each membrane, we use the AFM to indent the centre of suspended area with a force $F$ while measuring the cantilever indentation $\delta$ \cite{castellanos2012elastic}. The measured $F$ versus $\delta$, as depicted in Fig.~\ref{fig:1}b for device D1, is fitted with a model for point-force loading of a circular plate given by $F = (\frac{16\pi D}{R^2}\delta ) + n_0\pi\delta+Ehq^3(\frac{\delta^3}{R^2})$, where $D = Eh^3/(12(1-\nu^2))$ is the bending rigidity of the membrane, $\nu$ is Poisson ratio, $n_0=Eh\epsilon_0/(1-\nu)$ is the initial tension in the membrane, and $\epsilon_0$ is the prestrain. We extract $E=108.45$~\si{GPa} and $\epsilon_0 = 6.75\times 10^{-3}$ from the fit shown in Fig.~\ref{fig:1}b (drawn line), which are in good agreement with typical values found in literature for 2H-TaS$_2$ \cite{lee2021study}. The obtained values of $E$ for devices D2$-$D5 are listed in Table~\ref{tab:op_table1}.  

\begin{table*}
\caption{\label{tab:op_table1}
Characteristics of devices D1$-$D5, including radius $R$, thickness $h$, mass density $\rho$, Young's modulus $E$, atomic mass $M$, Poisson ratio $\nu$, Gr$\ddot{\text{u}}$neisen parameter $\gamma$, as well as the obtained average TEC $\alpha_m$, specific heat $C_v$ and in-plane thermal conductivity $k$. The values of $\rho$, $M$, $\nu$ and $\gamma$ are taken from literature \cite{lee2021study, kargar2020phonon, uma2001temperature, bae2017thickness,kumar2015thermoelectric, vsivskins2020magnetic}.} 

\begin{tabular}{lcccccccccc}
  \hline\hline
 & $R$ (\si{\mu m}) & $h$ (\si{nm}) & $\rho$ (\si{kg~m^{-3}}) & $E$ (\si{GPa}) & $M$ (\si{g~mol^{-1}}) & $\nu$ & $\gamma$ &  $\alpha_m$ ($\times 10^{-6} $ \si{K^{-1}}) & $C_v$ (\si{J mol^{-1} K^{-1}}) & $k$ (\si{W m^{-1} K^{-1}}) \\
\hline
D1 (2H-TaS$_2$) & 4 & 23.2 & 6860 & 108.45 & 245 & 0.35 & 2.13 & 6.96 & 42.0  & 8.6
\\
D2 (FePS$_3$) & 4 & 33.9 & 3375 & 69.60 & 183 & 0.304 & 1.80 &  12.7 & 68.2 & 1.8
\\
D3 (Poly Si) & 4 &  24.0 & 2330 &  140.52 & 28  &  0.22 & 0.45 & 3.10 & 20.7 & 5.3
 \\
D4 (MoS$_2$) & 3 & 4.8 & 5060 & 174.32 &  160 & 0.25 & 0.41 & 3.37 & 90.6 & 28.8
\\
D5 (WSe$_2$) & 3 & 5.5 & 9320 & 94.42 & 342 & 0.19 & 0.79 & 7.63 & 53.8 & 11.0
\\
  \hline\hline
\end{tabular}
\end{table*}

The setup for the optomechanical measurements \cite{Siskins2021tunable, liu2023tuning}, is shown in Fig.~\ref{fig:1}c. A power-modulated blue diode laser ($\lambda=405~$\si{nm}) photothermally actuates the resonator, while a He-Ne laser ($\lambda=632$~\si{nm}), of which the reflected laser power depends on the position of the membrane, is used to detect the motion of the resonator. The power-modulation of the blue laser is supplied by a Vector Network Analyzer (VNA), which also analyzes the photodiode signal containing the reflected laser power and converts that to the response amplitude, $|z_f|$, of the resonator in the frequency domain. All measurements were done in vacuum at a pressure below 10$^{-5}~\si{mbar}$. As shown in Fig.~\ref{fig:1}d, $|z_f|$ shows a clear fundamental resonance peak, which we fit with a harmonic oscillator model, given by $|z_{f}| = \frac{A_{\text{res}}f_0^2}{Q\sqrt{(f_0^2-f^2)^2+(f_0f/Q)^2}}$, where $f_0$ is the fundamental resonance frequency, $A_{\text{res}}$ is the vibration amplitude at resonance and $Q$ is the quality factor. For device D1, we obtain a $f_0=9.53$~\si{MHz} and $Q=160$. In addition, we also find a maximum in the imaginary part of $z_f$ at kHz frequencies (see Fig.~\ref{fig:1}e), which we attribute to the thermal expansion of the membrane that is time-delayed with respect to the modulated blue laser power, because it takes a time $\tau$ for the temperature of the membrane to rise \cite{dolleman2018transient,liu2022tension}. By solving the in-plane heat equation in the membrane, the thermal signal can be expressed as: 
\begin{equation}
    z_f = \frac{A_{\text{th}}}{i2\pi f \tau+1},
   \label{eq:1}
\end{equation}
where $A_{\text{th}}$ and $\tau$ are the thermal expansion amplitude and thermal time constant of the membrane, respectively. The red and blue laser powers are fixed at 0.9 and 0.13~\si{mW} respectively, to ensure linear vibration of the resonators with a negligible temperature raise of the membrane due to self-heating \cite{dolleman2018transient}. We extract $\tau$ by fitting the measured imaginary part of $z_f$ to Eq.~(\ref{eq:1}) (see Fig.~\ref{fig:1}e). Here, we obtain the maximum of $\text{Im}(z_f)$ at around 366.19~\si{kHz} for device D1, corresponding to $\tau=(2\pi\times366.19$~\si{kHz}$)^{-1}=434.62$~\si{ns}. 

\section{Results}

\subsection{Thermally-induced buckling phenomenon}

\begin{figure*}
	\centering	\includegraphics[width=1\linewidth,angle=0]{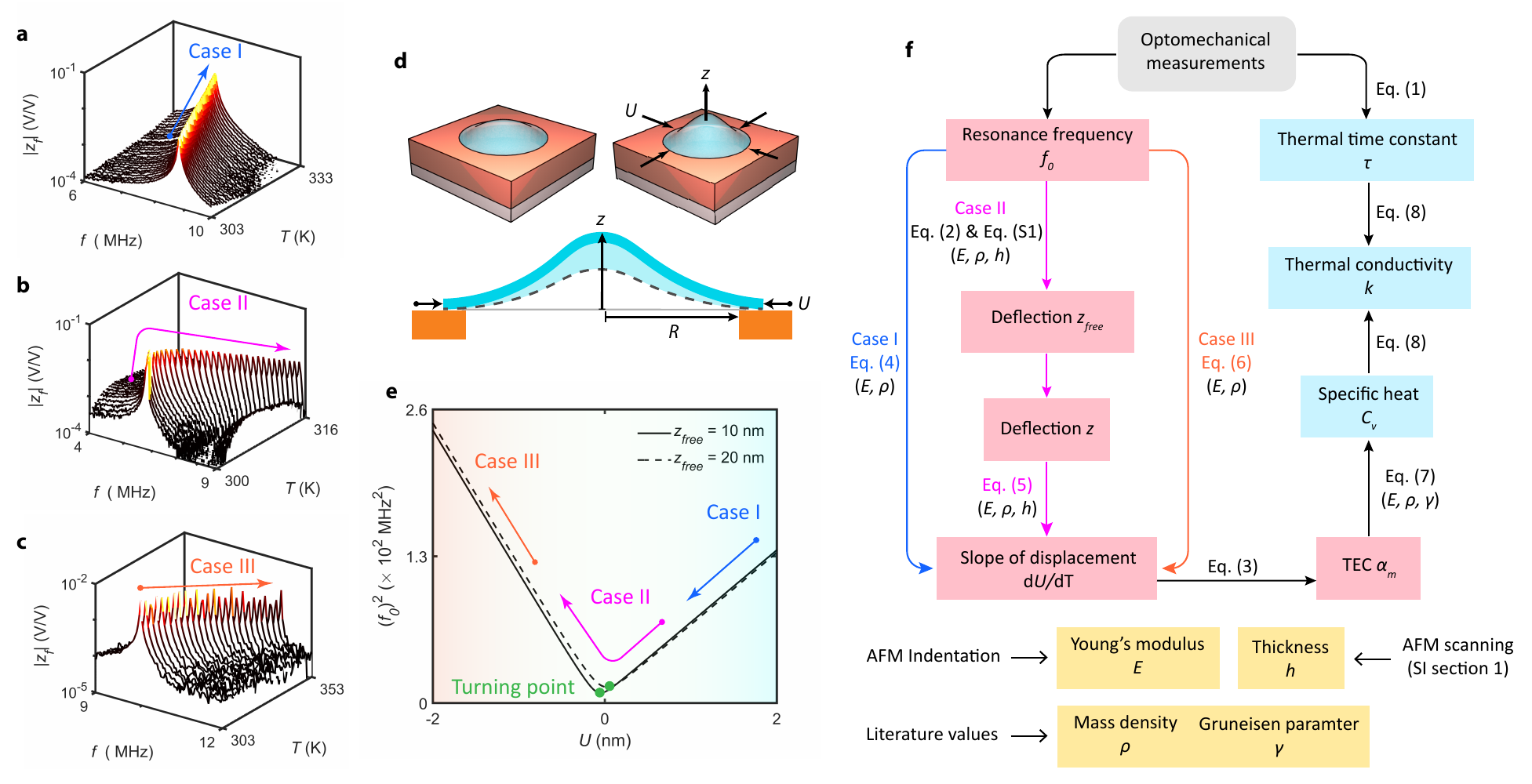}
	\caption{Optomechanical methodology for obtaining thermal properties. \textbf{a}-\textbf{c} Measured resonant peak measured as the function of temperature $T$ for devices D1 to D3, corresponding to cases I to III, respectively. \textbf{d} Schematic diagram of the buckled device, where the central deflection $z$ of the membrane increases as the boundary displacement $U$ loaded. \textbf{e} Squared resonance frequency $f_0^2$ as the function of $U$ in the membrane estimated by Eq.~(\ref{eq.T-frequnecy}) and Eq.~(S1) under different $z_{free}$. $f_0^2$ first decreases and then increases again as $U$ varies from tensile to compressive, which is comparable to the measurement result for device D2. \textbf{f} The proposed procedure to determine thermal properties of 2D material membranes.    
    }
	\label{fig:2}
\end{figure*}

When changing the temperature, the thermal expansion coefficient (TEC) $\alpha_m$ of the membrane, which is higher than that of the silicon substrate $\alpha_{\text{Si}}$, changes the strain in the membrane by a quantity $\Delta\epsilon$. This results in a remarkable change in the dynamics of 2D nanomechanical resonators, which can be used for probing the thermal properties \cite{ye2018electrothermally, liu2015optical}. Therefore, we heat up the fabricated devices and investigate the dependence of resonance frequency $f_0$ on temperature $T$. As shown in Fig.~\ref{fig:2}a, we observe a decrease of $f_0$ with increasing $T$ for device D1, which is in agreement with trends shown in literature \cite{wang2021thermal} and can be attributed to a reduction in strain when the material thermally expands. However, the results obtained for devices D2 and D3 are substantially different: as shown in Figs.~\ref{fig:2}b and \ref{fig:2}c, we observe an initial decrease in $f_0$ with increasing $T$ towards a minimum frequency (which we call the turning point), followed by a continuous increase. We attribute this to the thermally-induced buckling of the mechanical resonators as found in earlier studies \cite{kim2021buckling,rechnitz2022dc,kanj2022ultra}, which is caused by the loaded compression since $\alpha_m > \alpha_{\text{Si}}$. Here, as depicted in Fig.~\ref{fig:2}d, the thermal expansion of the membrane causes a compressive stress that triggers the membrane to buckle. We label the pre-buckling, the transition from pre- to post-buckling, and the post-buckling regions in Fig.~\ref{fig:2}e as cases I, II and III, respectively. 

We use a Galerkin model for a clamped circular plate \cite{yamaki1981non,kim1986flexural}, to find an approximate analytical expression of the fundamental resonance frequency $f_0$ under thermally-induced buckling \cite{liu2023enhanced}:  
\begin{widetext}
\begin{equation}
f_0(T) =  \frac{10.33h}{\pi d^2}\sqrt{\frac{E}{3\rho (1-\nu^2)} \left(1 + \beta(1-\nu^2)\frac{3z^2-z_{free}^2}{h^2} + \frac{3}{8}(1+\nu)\frac{Ud}{h^2} \right)},
    \label{eq.T-frequnecy}
\end{equation}
\end{widetext}
where $d$ is the diameter of the plate, $U$ is the thermally changed in-plane displacement from boundary, $\rho$ is the mass density, $z$ is the central deflection of the plate, $\nu$ is the Poisson ratio, $z_{free}$ is the central deflection of the plate in the pre-deformed state when $U=0$ (without loading), and $\beta $ is a fitting factor determined by \cite{liu2023enhanced} $\beta = 0.35\nu + 0.42$. Equation~(\ref{eq.T-frequnecy}) shows that $f_0$ depends on the in-plane displacement $U(T)$ from boundary and the central deflection $z(T)$ of the membrane. The relation between $z(T)$ and $U(T)$ can be found in SI section 1 (see Eq.~(S1)). Following the literature \cite{liu2023enhanced}, $z_{free}$ can be extracted from the measured value of the fundamental resonance frequency at the turning point, $f_t$, using the built Galerkin model.

By substituting $R=4$~\si{\mu m}, $h = 23$~\si{nm}, $E=108$~\si{GPa}, $\rho = 6860$~\si{kg /m^{3}}, $\nu=0.35$ and $\beta = 0.54$ into Eq.~(\ref{eq.T-frequnecy}) and Eq.~(S1), we obtain $f_0^2$ versus $U$ as shown in Fig.~\ref{fig:2}e. For case I, $f_0^2$ decreases as $U$ increases; while for cases cases II and III, $z$ increases as buckling happens (see Fig.~S2b), leading to an increase of $f_0^2$. 
The estimation in Fig.~\ref{fig:2}e can thus account for all measured results of $f_0$ versus $T$ for devices D1 to D3. In the following, we describe how to extract the slope of thermal-changed displacement $U$ versus temperature $\frac{\text{d}U}{\text{d}T}$ for cases I to III, which is related to the TEC $\alpha_m$ of the membrane through \cite{ye2018electrothermally}:
\begin{equation}
    \frac{1}{R}\frac{\text{d}U}{\text{d}T} = - [\alpha_m(T)-\alpha_{\text{Si}}(T)].
   \label{eq:TEC calculation}
\end{equation}

\renewcommand{\thesubsubsection}{Case \Roman{subsubsection}} 
\subsubsection{Pre-buckling regime}

For case I, the suspended membrane is nearly flat while the change of deflection $z$ with increasing temperature $T$ can be negligible. Therefore, assume $\frac{\text{d}z}{\text{d}T} = 0$, the derivative of Eq.~(\ref{eq.T-frequnecy}) can be simplified as (see details in SI section 2):
\begin{equation}
    \frac{\text{d}f_0^2}{\text{d}T } = c_t\frac{\text{d}U}{\text{d}T }.
   \label{eq:case I}
\end{equation}
where $c_t = \frac{13.34E}{\pi^2d^3\rho(1-\nu)}$. Therefore, in the pre-buckling regime, we can directly extract $\frac{\text{d}U}{\text{d}T}$ from the measured $\frac{\text{d}f_0^2}{\text{d}T}$ using Eq.~(\ref{eq:case I}) (see the flow chart in Fig.~\ref{fig:2}f). Besides device D1, we also show that devices D4 and D5 are in case I according to their measured $f_0$ versus $T$ (see SI section 4).

\renewcommand{\thesubsubsection}{Case \Roman{subsubsection}} 
\subsubsection{Transition from pre- to post-buckling}

For case II, Eq.~(\ref{eq:case I}) is not applicable anymore since $z(T)$ varies with temperature. We thus calculate the derivative of Eq.~(\ref{eq.T-frequnecy}) (see SI section 2) and obtain:
\begin{equation}
    \frac{\text{d}f_0^2}{\text{d}T } =  c_t\left(1-\frac{32}{\frac{16}{3\beta(1-\nu^2)}\frac{z_{free}h^2}{z^3} + 10.7} \right) \frac{\text{d}U}{\text{d}T },
   \label{eq:case II}
\end{equation}
As depicted in the flow chart of Fig.~\ref{fig:2}f, we first extract $z_{free} = 20.6$~\si{nm} for device D2 from the measured $f_0$ at the turning point using Eq.~(\ref{eq.T-frequnecy}) and Eq.~(S1), as well as $z$ versus $T$ (see Fig.~S2). The obtained $z_{free}$ and $z(T)$ are then substituted into Eq.~(\ref{eq:case II}) to extract $\frac{\text{d}U}{\text{d}T}$. The result of $U$ versus $T$ for device D2 shows the expected transition of displacement from tensile ($U>0$) to compressive ($U<0$), as plotted in Fig.~S2. 

\renewcommand{\thesubsubsection}{Case \Roman{subsubsection}} 
\subsubsection{Post-buckling regime}

For case III, Eq.~(\ref{eq:case II}) can be simplified as $z^3 \gg z_{free}h^2$, which results in:
\begin{equation}
     \frac{\text{d}f_0^2}{\text{d}T } =  -2c_t \frac{\text{d}U}{\text{d}T }.
   \label{eq:case III}
\end{equation}
Thus the result of $\frac{\text{d}U}{\text{d}T}$ for device D3 can be directly extracted from the measured $f_0$ versus $T$. The calculated curves in Fig.~\ref{fig:2}e also verify the linear relations given by Eq.~(\ref{eq:case I}) and Eq.~(\ref{eq:case III}).

\subsection{Extracting in-plane thermal conductivity of 2D materials}

The flow chart depicted in Fig.~\ref{fig:2}f also shows how optomechanical measurements as a function of temperature enable a precise pathway for studying the thermal properties of 2D resonators. We first extract the TEC $\alpha_m$ of the membrane from the the results of $\frac{\text{d}U}{\text{d}T}$ for cases I to III, which are obtained from Eqs.~(\ref{eq:case I}) to (\ref{eq:case III}), respectively, as discussed in the previous section. Then, we quantify the specific heat $c_v$ of the membrane from its thermodynamic relation with $\alpha_m$, which will be discussed in this section in more detail. Finally, from the solution of the 2D heat equation, we determine the in-plane thermal conductivity $k$ of the membrane from the measured $\tau$ and the obtained $c_v$. In the following, we go step by step through this procedure for device D1. 

\begin{figure*}
	\centering
\includegraphics[width=0.85\linewidth,angle=0]{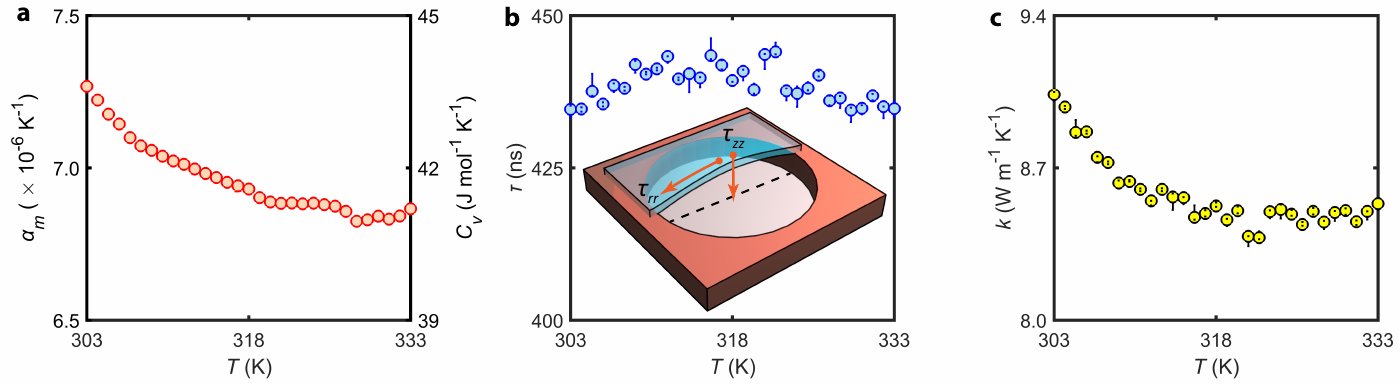}
	\caption{Quantifying thermal characteristics of device D1. \textbf{a} Thermal expansion coefficient $\alpha_m$ and specific heat $C_v$ of 2H-TaS$_2$ membrane versus temperature $T$. \textbf{b} Measured thermal time constant $\tau$ versus $T$. Insert, schematic diagram of heat transport in suspended 2D membrane. \textbf{c} In-plane thermal conductivity $k$ versus $T$ extract from Eq.~(\ref{eq: k versus T}) using the measured $\tau$ and the obtained $C_v$. }
	\label{fig:3}
\end{figure*}

Let us start with extracting the TEC $\alpha_m$ of the membrane. Since the in-plane displacement $U$ originates from the boundary thermal expansion of the membrane, we can extract the TEC $\alpha_m(T)$ of the membrane from the obtained $\frac{\text{d}U}{\text{d}T}$ using Eq.~(\ref{eq:TEC calculation}), where the values of $\alpha_\text{Si}(T)$ are taken from literature \cite{okada1984precise}. The obtained $\alpha_m$ versus $T$ for device D1 is shown in Fig.~\ref{fig:3}a (left).

In the second step, since the specific heat at constant volume is approximately equal to that at constant pressure for solid, we can directly extract the specific heat, $C_v$, of the membrane from the TEC $\alpha_m$ using the thermodynamic relation \cite{vsivskins2020magnetic}:
\begin{equation}
    C_v=\frac{3\alpha_m K V_{\text{M}}}{\gamma\rho},
   \label{eq: thermodynamic relation}
\end{equation}
where $K=\frac{E}{3(1-2\nu)}$ is the bulk modulus, $V_{\text{M}}=M/\rho$ is the molar volume, $M$ is the atomic mass, and $\gamma$ is the Gr$\ddot{\text{u}}$neisen parameter of the membrane taken from literature. These parameters are listed in Table~\ref{tab:op_table1} for the used materials. Using the obtained $\alpha_m$, we extract $C_v$ versus $T$ for device D1, as plotted in Fig.~\ref{fig:3}a (right).  

In the last step, we focus on the heat transport in 2D membranes. As shown in Fig.~\ref{fig:3}b, we experimentally observe that $\tau$ is between 434.6 and 444.0~\si{ns} in the probed $T$ range for device D1. Considering the heat transport in a circular membrane, we solve the heat equation in the membrane with an appropriate initial temperature distribution and well-defined boundary conditions (see SI section 3), and obtain the thermal time constant based on the thermal properties of the membrane:
\begin{equation}
     \tau^{-1} = {\tau_{rr}}^{-1} + {\tau_{zz}}^{-1} = \frac{k}{\rho c_v} \left( \frac{\mu^2}{R^2} + \frac{\pi^2}{4h^2}\right ),
   \label{eq: k versus T}
\end{equation}
where $\tau_{rr}$ and $\tau_{zz}$ are the in-plane and out-of-plane thermal time constants of the membrane (see Fig.~\ref{fig:3}b, insert), respectively, $c_v = C_v/M$, $\mu^2=5$ is the in-plane diffusive constant (see SI section 3), and $k$ is the thermal conductivity of the membrane. Due to the low $h/R$ ratio for 2D materials, we find that $\tau_{zz}\ll \tau_{rr}$ and thus the extracted $\tau$ from our measurement is equal to $\tau_{rr}$. By substituting the obtained $C_v$ and the measured $\tau$ into Eq.~(\ref{eq: k versus T}), we extract $k= 8.6 \pm 0.3$~\si{W m^{-1} K^{-1}} for device D1, as plotted in Fig.~\ref{fig:3}c. 

\begin{table*}
\caption{\label{tab:method comparison}
Comparison of different thermal conductivity measurement methods, where the required temperature range is quantified by the studies of MoS$_2$.} 
\begin{tabular}{llll}
  \hline\hline
 & Raman microscopy & Micro-bridge method & Optomechanics \\
\hline
Required temperature range & 50$-$100~\si{K} \cite{zhang2015measurement, sahoo2013temperature} &  10$-$50~\si{K} \cite{jo2014basal}  & $<10$~\si{K}
\\
Sample preparation &  Easy  & Difficult  &  Easy
\\
Applicability to 2D materials &  Applicable & Limited  & Applicable 
\\
  \hline\hline
\end{tabular}
\end{table*}

\begin{figure}
	\centering
	\includegraphics[width=0.9\linewidth,angle=0]{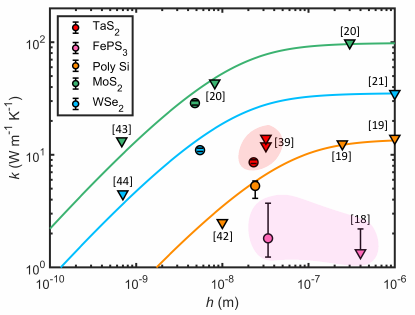}
	\caption{In-plane thermal conductivity $k$ of 2D material membranes versus their thickness $h$. Circular points, the obtained $k$ for devices D1$-$D5 in this work; triangle points, literature values; drawn lines, estimations of $k$ versus $h$ by Fuchs–Sondheimer model using Eq.~(\ref{eq:t versus k}).       
    }
	\label{fig:4}
\end{figure}

The obtained in-plane thermal conductivity $k$ for all devices D1$-$D5 are listed in Table~\ref{tab:op_table1}, of which the raw data can be found in Fig.~S6. For both 2H-TaS$_2$ (device D1) and FePS$_3$ (device D2), since relevant studies on their thermal properties are quite limited, we directly compare the obtained $k$ with the values from the literature \cite{kargar2020phonon, ccakirouglu2020thermal} and observe good agreements (see Fig.~\ref{fig:4}). For Poly Si (device D3), MoS$_2$ (device D4) and WSe$_2$ (device D5), we observe that $k$ depends on the membrane thickness $h$. We attribute this to a smaller mean free path (MFP) of phonons in thin membranes compared to their bulk counterparts \cite{luo2015anisotropic}. To account for this effect, we use the Fuchs–Sondheimer model \cite{sondheimer2001mean,bae2017thickness} that evaluates thermal conductivity of 2D materials as a function of thickness:  
\begin{equation}
     \frac{k}{k_{\text{bulk}}}\approx  1 - \frac{3}{8}\frac{\Lambda_{\text{bulk}}}{h} 
+ \frac{3}{2}\frac{\Lambda_{\text{bulk}}}{h}\int_1^{\infty} \left ( \frac{1}{x^3} - \frac{1}{x^5} \right ) e^{-\frac{h}{\Lambda_{\text{bulk}}}x} \text{d}x,
   \label{eq:t versus k}
\end{equation}
where $k_{\text{bulk}}$ and $\Lambda_{\text{bulk}}$ are the thermal conductivity and MFP of bulk, and $x$ is a integration variable. The bulk thermal conductivities $k_{\text{bulk}}$ for Poly Si, MoS$_2$ and WSe$_2$ are 13.8~\si{W m^{-1} K^{-1}}, 98.5~\si{W m^{-1} K^{-1}}, and 35.3~\si{W m^{-1} K^{-1}}, respectively \cite{liu2014measurement, uma2001temperature, kumar2015thermoelectric}. We find that the given $k$ versus $h$, including our results and literature values \cite{kargar2020phonon, uma2001temperature, bae2017thickness, ccakirouglu2020thermal, kumar2015thermoelectric, braun2016size, arrighi2021heat, norouzzadeh2017thermal}, is well described by Eq.~(\ref{eq:t versus k}) as indicated by the fitted solid lines in Fig.~\ref{fig:4} using $\Lambda_{\text{bulk}}$ as fit parameter, obtaining 75~\si{nm}, 19~\si{nm}, and 19~\si{nm} for Poly Si, MoS$_2$ and WSe$_2$, respectively. These fitted values of $\Lambda_{\text{bulk}}$ are also in good agreement with previously reported phonon MFPs \cite{uma2001temperature, liu2013phonon, cui2014transient}, supporting the validity of employing Eq.~(\ref{eq:t versus k}) to predict the thickness dependent thermal conductivity of 2D materials.

\section{Discussion}

Compared to other methods for determining the thermal conductivity of 2D materials, the optomechanical approach has several advantages, as summarized in Table~\ref{tab:method comparison}. For the Raman microscopy method, since relatively large temperature changes are needed to resolve the resulting shift in Raman mode frequency, a very wide temperature range has to be measured to get an accurate slope $\chi_T$ of the Raman peak shift with temperature. For example, $\chi_T$ for MoS$_2$ is -0.013~\si{cm^{-1}/K}. Considering a limited resolution 0.25~\si{cm^{-1}} for a Raman microscope, a temperature increase of at least 20~\si{K} is required to obtain meaningful results \cite{kasirga2020thermal}. In our measurements, we require only a narrow $T$ range to study the thermal transport (see Fig.~S6). For the micro-bridge method, either thick crystals or stiff 2D materials like graphene are required to survive the complicated fabrication procedures including lithography and etching \cite{wang2017thermal}. In contrast, for the presented contactless optomechanical method, one only needs to suspend membranes over cavities in a Si substrate, which is applicable for most 2D materials and can be done for any thickness.

Although we estimate the average MFP for bulk in Fig.~\ref{fig:4}, we note that the phonon MFP in 2D materials is highly related to the phonon dispersion relation, surface strain, crystal grain size, and temperature. These factors can be further studied using the presented optomechanical approach, which would help us to better understand the phonon scattering mechanisms in 2D materials. Moreover, our work suggests a new way to further investigate acoustic phonon transport in recently emerged 2D materials, such as phosphorene and MXenes with distinct thermal anisotropy \cite{qin2018thermal,frey2019surface}, as well as the magic-angle multilayer superconductor family \cite{park2022robust}. It is also of interest to probe the dynamics of phonons across the interface in vdW heterostructures, so as to realize a coherent control of thermal transport across 2D interfaces \cite{wu2021phonon,ren2021impact}.   

\section{Conclusions}

We demonstrated an optomechanical approach for probing the thermal transport in 2D nanomechanical resonators made of few-layer 2H-TaS$_2$, FePS$_3$, Poly Si, MoS$_2$, and WSe$_2$. We measured the resonance frequency and thermal time constant of the devices as a function of temperature, which are used to extract their thermal expansion coefficient, specific heat, as well as in-plane thermal conductivity. The obtained values of all these parameters (see Table~\ref{tab:op_table1}) are in good agreement with values reported in the literature. Compared to other methods for characterizing the thermal properties of 2D materials, the presented contactless optomechanical approach requires a smaller temperature range, allows for easy sample fabrication, and is applicable to any 2D material. This work not only advances the fundamental understanding of phonon transport in 2D materials, but potentially also enables studies into the use of strain engineering and heterostructures for controlling heat flow in 2D materials.

\section*{ASSOCIATED CONTENT}
See the supplementary information on the methodology and data related to the theory, experiment and simulation.

\section*{Notes}
The authors declare no competing financial interest.

\begin{acknowledgments}
P.G.S. and G.J.V. acknowledge support by the Dutch 4TU federation for the Plantenna project. H.S.J.v.d.Z. and P.G.S. acknowledge funding from the European Union’s Horizon 2020 research and innovation program under grant agreement no. 881603. H.L. acknowledges the financial support from China Scholarship Council. C.B.C acknowledges the financial support from the European Union (ERC AdG Mol-2D 788222), the Spanish MICIN (2D-HETEROS PID2020-117152RB-100, co-financed by FEDER, and Excellence Unit “María de Maeztu” CEX2019-000919-M), the Generalitat Valenciana (PROMETEO Program and PO FEDER Program, Ph.D fellowship) and the Advanced Materials program (supported by MCIN with funding from European Union NextGenerationEU (PRTR-C17.I1) and by Generalitat Valenciana).
\end{acknowledgments}

\bibliography{main}

\pagebreak
\onecolumngrid

\section*{Supplementary Information}

\subsection*{Section 1: Sample fabrication and characterization} 

\setcounter{figure}{0}
\begin{figure} [h]
	\centering
    \renewcommand{\thefigure}{S\arabic{figure}}
	\includegraphics[width=0.6\linewidth,angle=0]{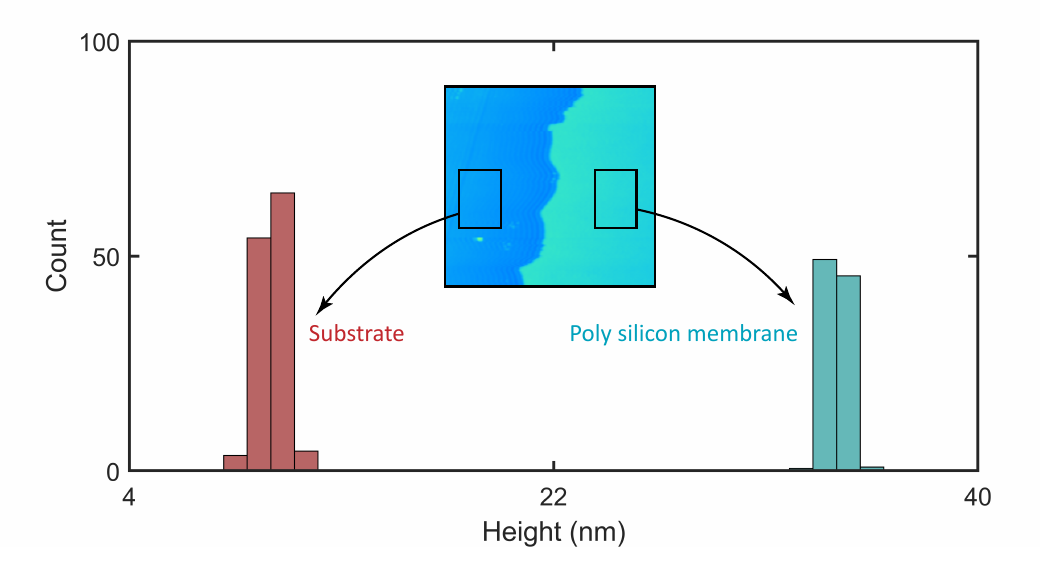}
	\caption{Height histogram of substrate (red), as well as Poly Si membrane (cyan), measured by tapping mode AFM. Insert, scanning image on the boundary of Poly Si membrane.}
	\label{fig:AFM of Pi silicon fabrication}
\end{figure}

A Si wafer with 285~\si{nm} dry SiO$_2$ is spin coated with positive e-beam resist and exposed by electron-beam lithography. Afterwards, the SiO$_2$ layer without protection is completely etched using CHF$_3$ and Ar plasma in a reactive ion etcher. The edges of cavities are examined to be well-defined by scanning electron microscopy (SEM) and AFM. After resist removal, 2D nanoflakes are exfoliated by Scotch tape, and then separately transferred onto the substrate at room temperature through a deterministic dry stamping technique. More details on the substrate fabrication and Scotach tape transfer method can be found in our previous work \cite{liu2022nanomechanical}. Using tapping mode atomic force microscopy (AFM), we measure the height difference between the membrane and the Si/SiO$_2$ substrate. As Fig.~\ref{fig:AFM of Pi silicon fabrication} shows, we find a membrane thickness $h$ of 24.0~\si{nm} for device D3.

The fabricated devices is then fixed on a sample holder inside the vacuum chamber. A PID heater and a temperature sensor are connected with the sample holder, which allows to precisely monitor and control the temperature sweeping. A piezo-electric actuator below the sample holder is used to optimize the in-plane XY position of the sample to maintain blue and red lasers irradiating on the center of the circular sample.

\subsection*{Section 2: Mechanical model of 2D material membranes under cases I to III} 

In the main text, we observe different types of dynamic response in the measured devices D1$-$D3, attributed to the thermally-induced buckling in 2D nanomechanical resonators. Eq.~(\ref{eq.T-frequnecy}) gives the expression of the resonance frequency $f_0(T)$ in buckled resonators. Accodrding to our previous work \cite{liu2023enhanced}, Eq.~(\ref{eq.T-frequnecy}) can be solved by the relation between central deflection $z$ of the membrane and compressive displacement $U$ from the boundary: 
\setcounter{equation}{0}
\begin{equation}
   \frac{32}{3}\left(1-\frac{z_{free}}{z}\right) - 10.7\beta(1-\nu^2)\left(\frac{z_{free}^2-z^2}{h^2} \right) + 4(1+\nu)\frac{Ud}{h^2} = 0,
   \tag{S1}
   \label{buckling deflection initial}
\end{equation}
where $U$ is the thermally induced in-plane displacement of the plate, $\rho$ is the mass density, $z$ is the central deflection of the plate, $\nu$ is the Poisson ratio, $z_{free}$ is the central deflection of free plate without loading (when $U=0$), and $\beta$ is a fitting factor determined by $\nu = 0.35\nu + 0.42$. Therefore, using Eq.~(\ref{buckling deflection initial}) and Eq.~(\ref{eq.T-frequnecy}), we can extract $z_{free} = 20.6$~\si{nm} and $z(T)$ from the measured $f_0(T)$ for FePS$_3$ device D2. As plotted in Fig.~\ref{fig:buckling of device D2}, we see as the boundary displacement $U$ varies from tensile to compressive with increasing $T$, the central deflection $z$ of the membrane gradually goes up. More details of the Galerkin buckling model can be found in our previous work \cite{liu2023enhanced}.  

For pre-buckling regime (case I), assume the deflection $z$ nearly keeps constant, the only time-dependent parameter in Eq.~(\ref{eq.T-frequnecy}) is $U$, which allows us to obtain the derivative $\frac{\text{d}f_0^2}{\text{d}T} = c_t \frac{\text{d}U}{\text{d}T}$ for case I. For transition regime from pre- to post-buckling (case II), we first calculate the derivative of Eq.~(\ref{buckling deflection initial}):
\setcounter{equation}{1}
\begin{equation}
  -\frac{\text{d}U}{\text{d}T} = \left [\frac{32z_{free}}{3z^2} + 10.7\beta (1-\nu^2)\frac{2z}{h^2}  \right ] \frac{h^2}{4d(1+\nu)} \frac{\text{d}z}{\text{d}T}.
  \tag{S2}
\label{buckling deflection derivative}
\end{equation}
Therefore, by substituting Eq.~(\ref{buckling deflection derivative}) into the $T$-derivative of Eq.~(\ref{eq.T-frequnecy}), we obtain Eq.~(\ref{eq:case II}). For post-buckling regime (case III), the thermally-induced buckling results in a large central deflection of the membrane, thus we assume $z_{free}h^2 \ll z^3$ and simplify Eq.~(\ref{eq:case II}) as:
\setcounter{equation}{3}
\begin{equation}
  \frac{\text{d}f_0^2}{\text{d}T} = c_t(1-\frac{32}{10.7}) \frac{\text{d}U}{\text{d}T} = -2c_t\frac{\text{d}U}{\text{d}T}.
  \tag{S3}
\label{buckling deflection post}
\end{equation}

\setcounter{figure}{1}
\begin{figure} [h]
	\centering
    \renewcommand{\thefigure}{S\arabic{figure}}
	\includegraphics[width=0.9\linewidth,angle=0]{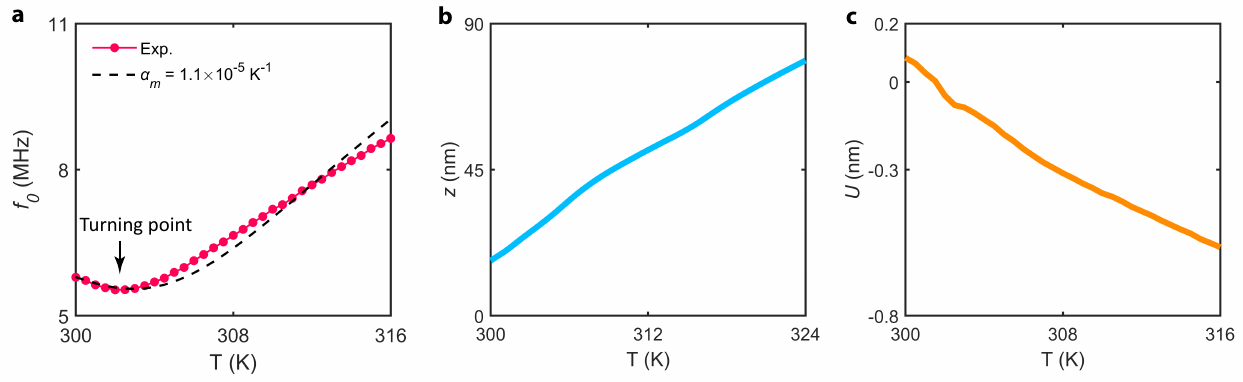}
	\caption{ Analysis of thermally-induced buckling for device D2. (a) Points, measured fundamental resonance frequency $f_0$ versus temperature $T$; drawn line, theoretical curve estimated by Eq.~(2) and Eq.~(\ref{buckling deflection initial}) using the TEC of FePS$_3$ $\alpha_m = 1.1 \times 10^{-5}$~\si{K^{-1}}. (b) Central deflection $z$ of the membrane versus $T$. (c) Boundary displacement of the membrane $U$ versus $T$.}
	\label{fig:buckling of device D2}
\end{figure}

\subsection*{Section 3: Heat transport model} 

In this section, we explain how we derive thermal time constant $\tau$ with respect to the thermal properties of 2D membrane, as given in Eq.~(8) in the main text.  

Consider the situation that a modulated-laser irradiates at the center of the suspended 2D membrane, the Fourier heat conduction equation in cylindrical coordinate for this problem can be written as:
\begin{equation}
   \frac{\partial u}{\partial t} = \kappa \left (  \frac{\partial^2 u}{\partial r^2} +\frac{1}{r}\frac{\partial u}{\partial r} + \frac{\partial^2 u}{\partial z^2}   \right ) + \frac{1}{c_p\rho} \frac{\text{d}Q}{\text{d}t} \quad \text{in}  \quad 0<z<h \quad 0<r<R,
   \label{eqs.heat equation}
   \tag{S4}
\end{equation}
where $u(r,z,t)$ is time $t$-dependent temperature distribution in the membrane along with the radial $r$ and perpendicular $z$ directions, $\kappa=\frac{k}{c_p\rho}$ is thermal diffusivity, $\frac{\text{d}Q}{\text{d}t}$ is the absorbed heat energy per unit time per unit volume at the center of the membrane, $c_p$ and $k$ denote the specific heat and thermal conductivity of the membrane, respectively. 

\subsubsection*{3.1 Transient state}

We firstly discuss the transient state of heat transport in the membrane, corresponding to a laser pulse irradiates. As a result, Eq.~(\ref{eqs.heat equation}) is rewritten as:
\begin{equation}
   \frac{\partial u}{\partial t} = \kappa \left (  \frac{\partial^2 u}{\partial r^2} +\frac{1}{r}\frac{\partial u}{\partial r} + \frac{\partial^2 u}{\partial z^2}     \right ) \quad \text{in}  \quad 0<z<h \quad 0<r<R,
   \label{eqs.heat equation homo}
   \tag{S5}
\end{equation}
with the boundary conditions:
\begin{equation}
    u = 0 \quad \text{at} \quad r=R, \quad u =0 \quad \text{at} \quad z=0 \quad \text{and} \quad \frac{\partial u}{\partial z} =0 \quad \text{at} \quad  z=h, 
   \label{eqs.heat equation boundary}
   \tag{S6}
\end{equation}
and the initial condition:
\begin{equation}
    u = u_0(r,z) \quad \text{for} \quad t=0 \quad \text{in} \quad 0\leq z\leq h  \quad \text{and} \quad 0\leq r\leq R,
   \label{eqs.heat equation IC}
   \tag{S7}
\end{equation}

We adopt the Fourier method (separation of variables) to solve Eq.~(\ref{eqs.heat equation homo}) combined by the boundary and initial conditions. Thus temperature distribution has a general expression as $u(r,z,t)=\psi(r)\chi(z)\Gamma(t)$, of which the independent cases, $\psi(r)$, $\chi(z)$ and $\Gamma(t)$ are separately given by:
\begin{equation}
    \begin{cases}
    \frac{1}{\psi} \left ( \frac{d^2 \psi}{dr^2} +\frac{1}{r}\frac{d\psi}{dr} \right) = -\eta^2, \quad \psi = 0 \quad \text{at} \quad r=R \\
    \frac{1}{\chi}\frac{d^2\chi}{dz^2} = -\gamma^2, \quad \chi = 0 \quad \text{at} \quad z=0 \quad \text{and} \quad \frac{\partial \chi}{\partial z} = 0 \quad \text{at} \quad z=h \\
    \frac{1}{\Gamma}\frac{d\Gamma}{dt}=-\lambda^2,
    \end{cases}
   \label{eqs.separate expressions} 
   \tag{S8}
\end{equation}
where the constants $\eta$, $\gamma$ and $\lambda$ are determined by solving the above equations combined with the corresponding conditions. We derive the eigenvalue solution of the first term in Eq.~(\ref{eqs.separate expressions}) as:
\begin{equation}
   \psi_m(\eta_m,r) = J_0(\eta_m r) \quad , \quad m = 1,2,3,..,.
   \label{eqs.solution in r}
   \tag{S9}
\end{equation}
where $\eta_m$ is the $m$-th root if the formula $J_0(\eta_mR)=0$, e.g., $\eta_1=\frac{2.4048}{R}$. Next, the solution of the second term in Eq.~(\ref{eqs.separate expressions}) is given as:
\begin{equation}
   \chi_n(z) = a_n \sin{\frac{(2n-1)\pi z}{2h}} \quad , \quad n = 1,2,3,...,
   \label{eqs.solution in z}
   \tag{S10}
\end{equation}
where $a_n$ is the constant of integration, $\gamma_n=\frac{(2n-1)\pi}{2h}$ is called the eigenvalues of the Sturm-Liouville problem. Then, the solution of the third term in Eq.~(\ref{eqs.separate expressions}) is given as:
\begin{equation}
   \Gamma_{mn}(t) = b_{mn}e^{-\lambda_{mn}^2t} \quad , \quad m = 1,2,3,..., \quad n = 1,2,3,...,
   \label{eqs.solution in t}
   \tag{S11}
\end{equation}
where $b_{mn}$ is the constant of integration, $\lambda_{mn}^2 =\kappa (\eta_m^2+\gamma_n^2)$. Totally, combine the solutions together, we have, from Eqs.~(\ref{eqs.solution in z}) to ~(\ref{eqs.solution in t}):
\begin{equation}
   u(r,z,t) = \sum_{m=1}^{\infty}\sum_{n=1}^{\infty}A_{mn}J_0(\eta_m r)(\sin{\frac{(2n-1)\pi z}{2h}})e^{-\lambda_{mn}^2t},
   \tag{S12}
   \label{eqs.solution in total}
\end{equation}
where the Fourier coefficient, $A_{mn}$, can be further extracted from the initial condition as \cite{hancock20061}:
\begin{equation}
   A_{mn} = \frac{ \int_{0}^{R}\int_{0}^{h} r \psi_m(\eta_m,r) \chi_n(z) u_0(r,z)dzdr }{\int_{0}^{R}r \psi_m^2(\eta_m,r) dr \int_{0}^{h}\chi_n^2(z) dz }.
   \label{eqs.cofficient A_mn}
   \tag{S13}
\end{equation}
Finally, Eq.~(\ref{eqs.solution in total}) and Eq.~(\ref{eqs.cofficient A_mn}) are the derived solutions of the Heat Equation Eq.~(\ref{eqs.heat equation homo}) in the membrane, with the corresponding boundary conditions in Eq.~(\ref{eqs.heat equation boundary}).

Suppose the initial temperature distribution $u_0(r,z)$ in the membrane is constant, i.e. $u_0(r,z)=u_0$, we extract the series of Fourier coefficient as $A_{11}\approx2.04u_0$, $A_{12}\approx0.68u_0$, $A_{12}\approx0.41u_0$, $A_{31}\approx-0.62u_0$...; on the other hand, the values of constant $\lambda_{11}^2=\kappa (\eta_1^2+\gamma_1^2)$, $\lambda_{12}^2=\kappa (\eta_1^2+\gamma_2^2)$,..., making the ratio $\frac{\lambda_{mn}^2}{\lambda_{11}^2}\gg1$. This indicates that the first term in Eq.~(\ref{eqs.solution in total}) dominates the sum of the rest of the terms, allows us to approximately express the temperature distribution as:
\begin{equation}
   u(r,z,t)\approx 2.04u_0J_0(\frac{2.4048r}{R})(\sin{\frac{\pi z}{2h}})e^{-\lambda_{11}^2t}.
   \label{eqs.solution only 1st term}
   \tag{S14}
\end{equation}
As a result, the thermal time constant of the membrane, extracted from $\lambda_{11}^2$, can be expressed as:
\begin{equation}
   \tau_{rr} = \frac{1}{\kappa\eta_1^2} = \frac{R^2c_p\rho}{5.78k}, \quad \tau_{zz} = \frac{1}{\kappa\gamma_1^2} = \frac{4h^2c_p\rho}{\pi^2k},
   \label{eqs.tau}
   \tag{S15}
\end{equation}
where $\tau_{rr}$ and $\tau_{zz}$ represent the time constant along in-plane and across-plane directions, respectively. Note that the ratio $\frac{R}{h}$ is quite large in atomic-layer-thick 2D membrane, we thus have $\tau_{rr} \gg \tau_{zz}$.

\subsubsection*{3.2 Quasi-steady state}

Define the temperature distributions of transient and quasi-steady as $u_{trans}$ and $u_{quasi}$, respectively. It should be noticed that $u_{trans}$ will be negligible as the time $t\rightarrow\infty$, since $e^{-\lambda_{11}^2t}\rightarrow0$ in Eq.~(\ref{eqs.solution only 1st term}). As a result, we only focus on $u_{quasi}$ in the following and adopt $u=u_{quasi}$. In addition, to simplify discussion, we assume $u$ is uniform in $z$- direction in 2D membrane, and the heat transport here in the quasi-steady case converts to 1D problem. Hence, Eq.~(\ref{eqs.heat equation}) changes to:
\begin{equation}
\frac{\partial u}{\partial t} = \kappa \left ( \frac{\partial^2 u}{\partial r^2} + \frac{1}{r} \frac{\partial u}{\partial r} \right ) \quad \text{in}  \quad 0<z<h \quad 0<r<R.
   \label{eqs.1d heat}
   \tag{S16}
\end{equation}
Due to the optothermal drive, we consider the oscillatory boundary conditions (along $r$- axis):
\begin{equation}
u = A\cos(\omega t) \quad \text{at} \quad r=0 \quad \text{and} \quad  u =0 \quad \text{at} \quad r=R,
   \label{eqs.1d BC}
   \tag{S17}
\end{equation}
where $A$ denotes the amplitude of temperature changing in membrane's center and $\omega$ is the modulating angular frequency of the laser. As $t\rightarrow\infty$, the solution of $u$ will become periodic with $\omega$, i.e. $u(r,t)=A(r)\cos(\omega t+ \phi(r))$, where $A(r)$ and $\phi(r)$ are the amplitude and phase of the quasi-steady state, respectively. We rewrite $u(r,t)$ with complex form:
\begin{equation}
u(r,t) = \frac{1}{2}(U(r)e^{i\omega t}+U^{*}(r)e^{-i\omega t}),
   \label{eqs.complex u}
   \tag{S18}
\end{equation}
where $A(r)=|U(r)|$ and $\phi(r) = \arctan \left ( \frac{\text{Im}(U(r))}{\text{Re}(U(r))}\right )$. Substitute Eq.~(\ref{eqs.complex u}) back into Eq.~(\ref{eqs.1d heat}), we obtain:
\begin{equation}
    i\omega U(r)e^{i\omega t} - i\omega U^{*}(r)e^{-i\omega t} \\
            = \kappa \left ( U^{''}(r)e^{i\omega t}+U^{*''}(r)e^{-i\omega t} + \frac{1}{r} U^{'}(r)e^{i\omega t} + \frac{1}{r} U^{*'}(r)e^{-i\omega t} \right ) .
   \label{eqs.complex expansion}
   \tag{S19}
\end{equation}
Using the Lemma (zero sum of complex exponentials) condition, Eq.~(\ref{eqs.complex expansion}) can be simplified as:
\begin{equation}
i\omega U(r) - \kappa U^{''}(r) - \frac{\kappa}{r}U^{'}(r) = 0.
   \label{eqs.complex expansion simple}
   \tag{S20}
\end{equation}

According to Eq.~(\ref{eqs.complex expansion simple}), now the problem becomes to solve the formula:
\begin{equation}
\frac{1}{U(r)}\left( U^{''}(r) + \frac{1}{r}U^{'}(r) \right) = -\left( \sqrt{\frac{\omega}{2\kappa}}(1-i) \right)^2,
   \label{eqs.complex expansion simple final}
   \tag{S21}
\end{equation}
with the boundary conditions:
\begin{equation}
U(0)=A \quad \text{and} \quad U(R)=0.
   \label{eqs.complex expansion simple BC}
   \tag{S22}
\end{equation}
Note that Eq.~(\ref{eqs.complex expansion simple final}) has the same forms as the first terms of Eq.~(\ref{eqs.separate expressions}) in transient case. Solving it gives: 
\begin{equation}
U(r) = c_1J_0(mr) + c_2Y_0(mr),
   \label{eqs.complex expansion simple solution}
   \tag{S23}
\end{equation}
where $m=\sqrt{\frac{\omega}{2\kappa}}(1-i)$. The constants $c_1$ and $c_2$ are determined by the boundary conditions Eq.~(\ref{eqs.complex expansion simple BC}) as:
\begin{equation}
c_1 = \frac{AJ_0(mR)}{ Y_0(0)J_0(mR) -  Y_0(mR)J_0(0)} \quad \text{and} \quad  c_2 = \frac{AY_0(mR)}{ J_0(0)Y_0(mR)-Y_0(0)J_0(mR)}.  
   \label{eqs.complex expansion simple solution with BC}
   \tag{S24}
\end{equation} 
Finally, the average temperature $\overline{U}$ in the membrane can be expressed as:
\begin{equation}
\overline{U} = \frac{1}{R}\int_{0}^{R} U(r) dr.
   \label{eqs.average T}
   \tag{S25}
\end{equation}
Note that $Y_0(r)\rightarrow\infty$ as $r\rightarrow0$, we assign $Y_0(0)$ with a tiny $\Delta$ as $Y_0(\Delta)$, which then allows us to extract the effective and finite value of $\overline{U}$ integrated from Eq.(~\ref{eqs.average T}).

To verify the accuracy of the built model at quasi-steady state, we extract $\overline{U}$ as the function of laser driving frequency $f$ from Eq.~(\ref{eqs.average T}). Assume $R=4$~\si{\mu m}, $h=31.4$~\si{nm}, $\rho=3375$~\si{kg m^{-3}}, $c_p=700$~\si{J kg^{-1} K^{-1}}, and $k = 5$~\si{W m^{-1} K^{-1}} for the membrane, we obtain $\tau=1291.0$~\si{ns}, corresponding to a thermal diffusive constant $\mu^2 = 5.86$. This value is comparable to the obtained result of 5.78 at transient state.

\subsubsection*{3.3 COMSOL simulation}

We now calibrate the solution of heat equation using COMSOL simulation, as illustrated in Fig.~\ref{fig:simulation COMSOL}a. We first fix the radius of laser spot as its realistic value $R_0=0.25$~\si{\mu m}. While for the boundary condition, $u|_{r=R} = 0$, considering the substrate, we change it to the bottom of Si (see the insert in Fig.~\ref{fig:simulation COMSOL}a). The thicknesses of SiO$_2$ and Si layers are set at 285~\si{nm} and 1~\si{\mu m}, respectively, while the other parameters for the membrane, including $R$, $h$, $\rho$, $c_p$ and $k$, are used the given values in section 3.2. We obtain the simulated temperature distribution of the membrane, as shown in Fig.~\ref{fig:simulation COMSOL}b. By fitting Eq.~(1) to simulation, we extract $\tau = 1506.9$~\si{ns}, corresponding to $\mu^2 = 5.02$. This value is thus adopted in Eq.~(8) in the main text, allowing us to estimate in-plane thermal conductivity of all fabricated devices.  

\setcounter{figure}{2}
\begin{figure}
	\centering
 \renewcommand{\thefigure}{S\arabic{figure}}
	\includegraphics[width=1\linewidth,angle=0]{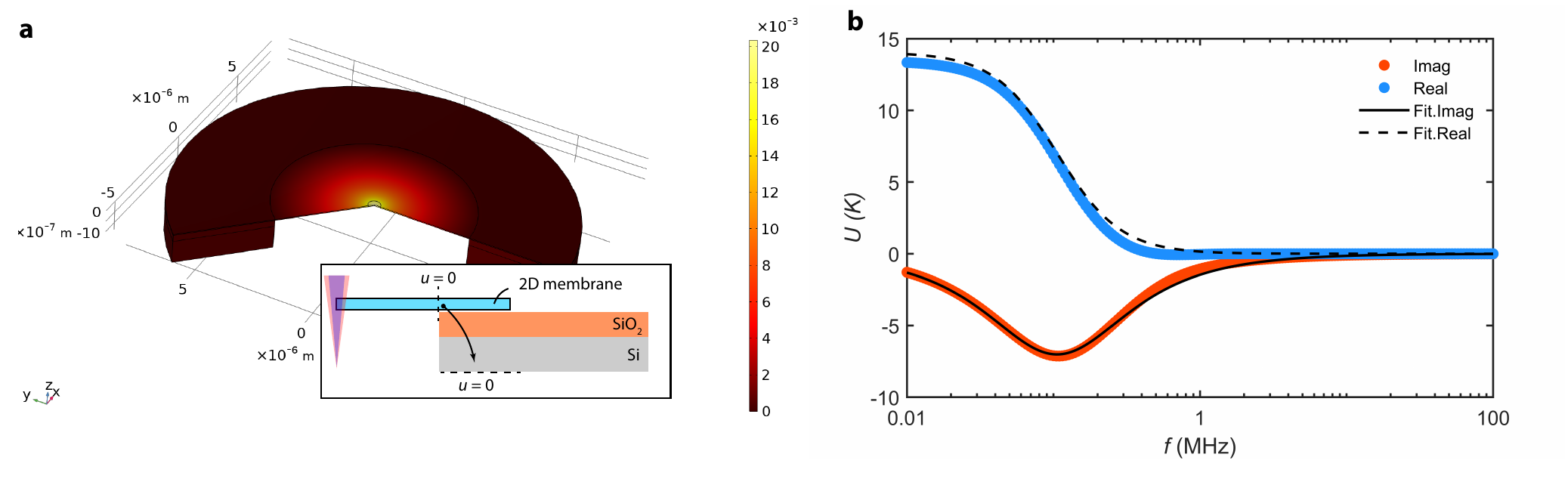}
	\caption{(a) 3D schematic diagram of 2D nanomechanical resonators in COMSOL simulation software. Insert, side view, where the boundary condition changes to the bottom of SiO$_2$/Si substrate. (b) Average temperature of the membrane as the function of heating rate $f$. Points, simulation results; lines, fit with Eq.~(1) in the main text to the simulation.}
	\label{fig:simulation COMSOL}
\end{figure}

\subsubsection*{3.4 Dependence on laser powers}

Furthermore, we discuss the effect of laser powers on our optomechanical measurement, as depicted in Fig.~\ref{fig:power discuss}. According to COMSOL simulation, we verify that the heating only plays a role on the amplitude of thermal signal, instead of its location which is related to $\tau$ (Fig.~\ref{fig:power discuss}a). Next, we test MoS$_2$ device D4, and plot the extract $\tau$ as the function of red and blue laser powers, respectively (see Figs.~\ref{fig:power discuss}b and \ref{fig:power discuss}c). As expected, the obtained $\tau$ nearly keep constant and are independent to both $P_{r}$ and $P_{b}$. Therefore, we verify that the proposed optomechanical methodology does not need a laser calibration for determining the in-plane thermal conductivity of 2D materials.   

\setcounter{figure}{3}
\begin{figure}
	\centering
 \renewcommand{\thefigure}{S\arabic{figure}}
	\includegraphics[width=0.9\linewidth,angle=0]{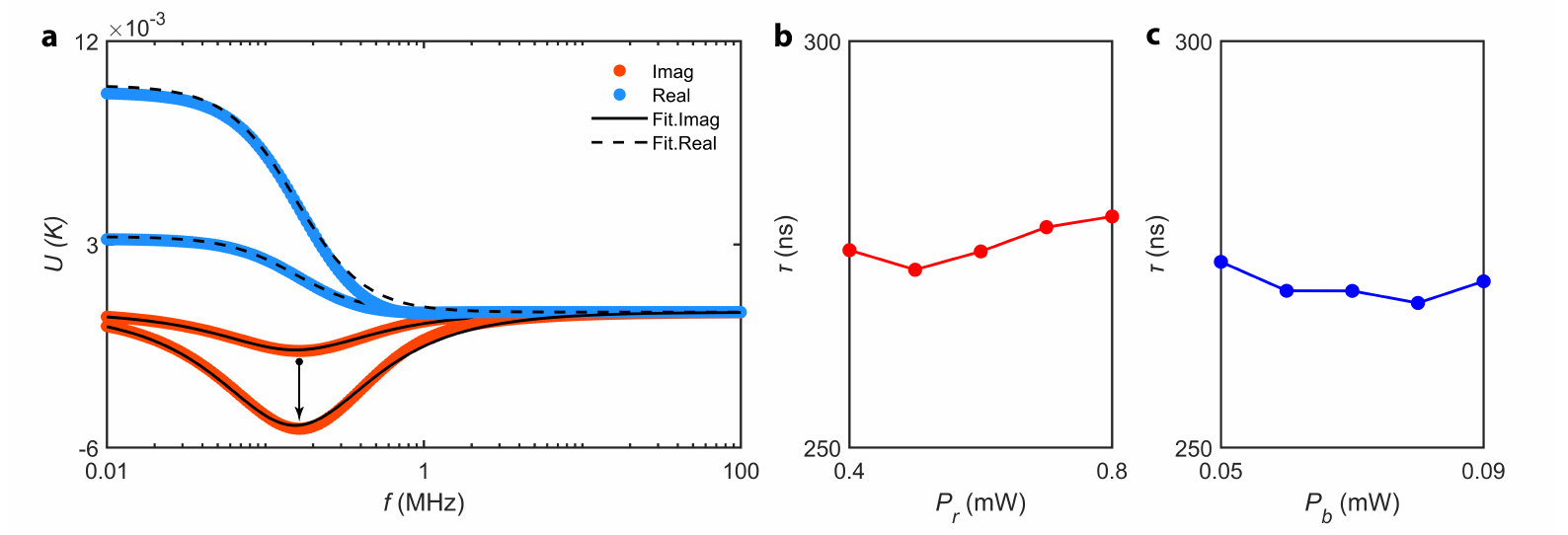}
	\caption{(a) COMSOL simulation results under different heating power. The amplitude of thermal signal increases as the heating power enhances from 1 to $3\times 10^{12}$~\si{W m^{-3}}, while its location is still fixed, indicating an unchanged thermal time constant $\tau$. The measured $\tau$ for a MoS$_2$ sample as the function of (b) red and (c) blue laser powers, respectively.}
	\label{fig:power discuss}
\end{figure}

\subsection*{Section 4: Raw data of optomechanical measurements for devices D2 to D5}

In Figs.~\ref{fig:f_0 devices D4 and D5}a and \ref{fig:f_0 devices D4 and D5}b, we see that the measured $f_0$ decreases as $T$ increases for devices D4 (MoS$_2$) and D5 (WSe$_2$). Therefore, both of them are in pre-buckling regime (case I), same as device D1 (2H-TaS$_2$). In addition, one could argue that the observed increase of $f_0$ with $T$ for device D3 (Fig.~2c in the main text) is attributed to $\alpha_m < \alpha_{\text{Si}}$, instead of a post-buckling performance. As shown in Fig.~\ref{fig:f_0 devices D4 and D5}c, we obtain the first decrease and then increase of $f_0$ with $T$ increasing (case II) for another measured Poly Si device, which verify that $\alpha_{\text{m}} > \alpha_{\text{Si}}$ for Poly Si and the observed increase of $f_0$ in device D3 corresponds to the mechanical response in post-buckling regime. 

\setcounter{figure}{4}
\begin{figure}
    \centering
     \renewcommand{\thefigure}{S\arabic{figure}}
    \includegraphics[width=1\linewidth,angle=0]{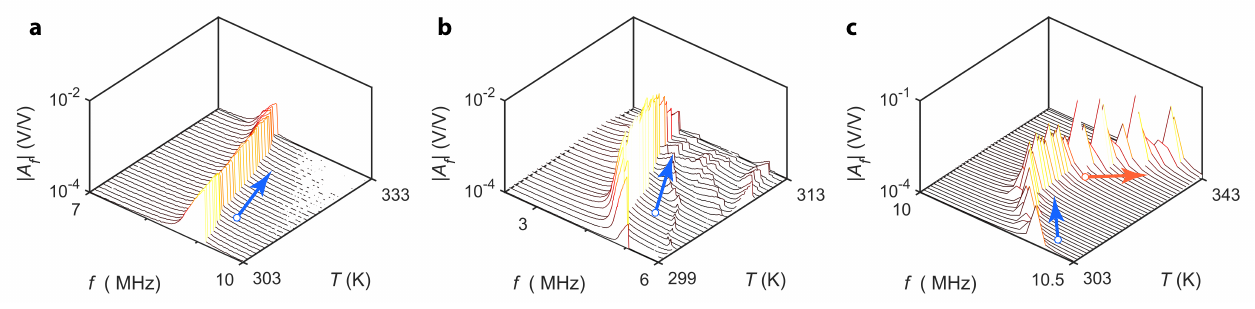}
    \caption{Measured resonance frequency $f_0$ as the function temperature $T$ for (a) MoS$_2$ device D4, (b) WSe$_2$ device D5, as well as (c) another Poly Si device with buckling transition.}
	\label{fig:f_0 devices D4 and D5}
\end{figure}  

Figure.~\ref{fig:all devices} shows the obtained specific heat $c_v$, the measured $\tau$ and the obtained in-plane thermal conductivity as the function of $T$ for the fabricated devices D2$-$D5, respectively. Corresponding results for 2H-TaS$_2$ device D1 has been given in Fig.~3 in the main text.

\setcounter{figure}{5}
\begin{figure}
	\centering
 \renewcommand{\thefigure}{S\arabic{figure}}
	\includegraphics[width=0.95\linewidth,angle=0]{"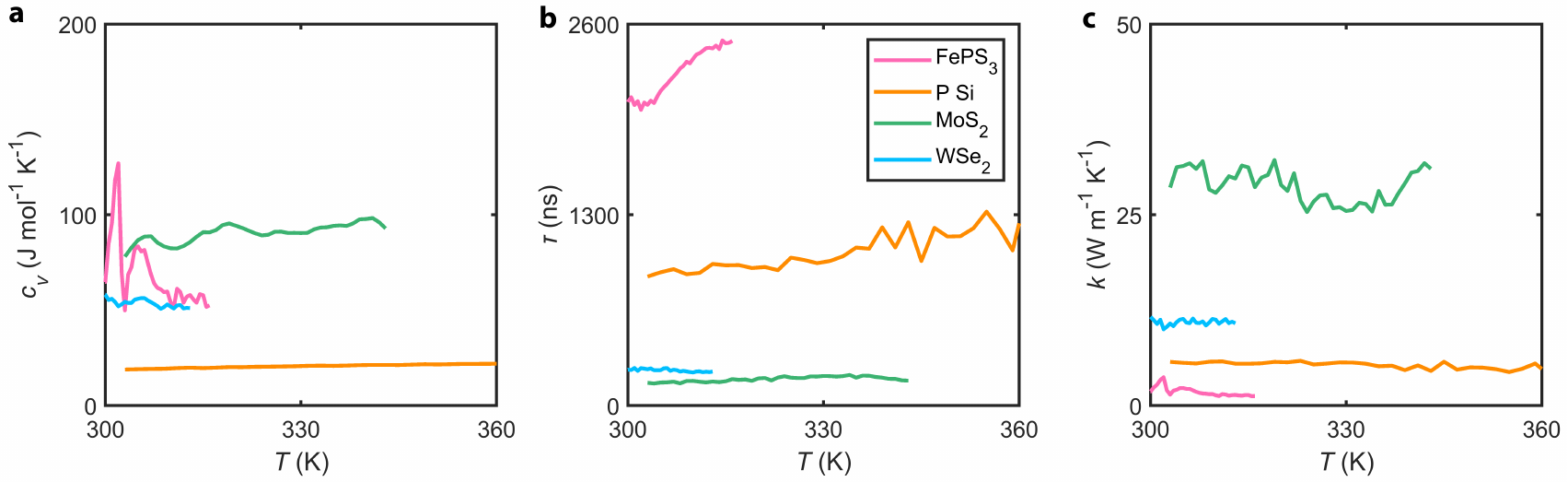"}
	\caption{(a)-(c) Specific heat $c_v$, thermal time constant $\tau$ and in-plane thermal conductivity $k$ as the function of $T$ for the fabricated devices D2$-$D5, respectively.}
	\label{fig:all devices}
\end{figure}

\end{document}